\documentclass[12pt]{iopart}

\usepackage{iopams}
\usepackage{amssymb}
\begin{document}

\title[Spontaneous emission of graviton by a quantum bouncer]{Spontaneous
emission of graviton by a quantum bouncer}

\author{G Pignol, K Protasov}

\address{Laboratoire de Physique Subatomique et de Cosmologie,
 CNRS/IN2P3-UJF-INPG, 53, Avenue des Martyrs, Grenoble, France}
\ead{pignol@lpsc.in2p3.fr}

\author{V Nesvizhevsky}
\address{Institute Laue Langevin, 6, rue Jules Horowitz, Grenoble, France}

\begin{abstract}
Spontaneous emission of graviton rates for the quantum bouncer
states are evaluated.
\end{abstract}



The quantum problem of a ball bouncing above a ideal mirror was
considered a long time ago, however just as a mere exercise of
elementary quantum mechanics. Things did change since the
quantization of energy of Ultra Cold Neutrons  (UCN) bouncing
above a mirror in the Earth's gravitational field had been
demonstrated in an experiment performed at the Institute Laue
Langevin (ILL) \cite{Nesvizhevsky:2002ef, Nesvizhevsky:2003ww,
Nesvizhevsky:2005}. However, this effect does not demonstrate any
quantum behavior of the gravitational field itself. In analogy
with electrodynamics, the observation of spectral lines in atoms
shows the quantum behavior of electrons, but does not provide any
clue concerning the possible quantization of the electromagnetic
field. What does provide a clue is the observation of spontaneous
decay of an excited state, for instance, which can only be
explained in terms of photon emission.

So the observation of spontaneous decay of an excited state in the
ILL experiment would be of interest, since it would be a
Planck-scale physics effect. Nevertheless, the decay rate is
expected to be low, and the purpose of this letter is to estimate
it.

First we will set notations of the quantum bouncer problem,
focusing on its physical implementation, that is, neutrons falling
on the Earth's gravitational field above a perfect mirror. We then
derive the spontaneous emission rate in a semi-classical approach.


The stationary Schr\"odinger equation for the vertical motion ($z$
axis) of the mass $m$ quantum bouncer is:
\begin{equation}
-\frac{\hbar^2}{2 m} \frac{d^2 \psi}{dz^2} + m g z \ \psi = E \
\psi. \label{schrodinger}
\end{equation}
The boundary condition due to the presence of the perfect mirror
at $z = 0$ is  $\psi(z = 0) = 0$. The characteristic length and
the characteristic energy of this problem are:
\begin{eqnarray}
z_0 & = \ \left( \frac{\hbar ^2}{2 m^2 g} \right)^{1/3} & = \ 5.87
\ \mu \rm{m} \\
E_0 & = \ \ m g z_0 & = \ 0.60 \ \rm{peV}.
\end{eqnarray}
The eigenproblem (\ref{schrodinger}) can be solved in terms of the
first Airy function $Ai(X)$, which has an infinite number of
negative zeros, denoted by $\{-\lambda_1, -\lambda_2, \dots\}$ in
decreasing order. The energy of the stationary states is equal to
$E_n = E_0 \lambda_n$, and the wave function of the $n^{th}$ state
is:
\begin{equation}
\psi_n(z) = C_n Ai \left(\frac{z}{z_0} - \lambda_n \right) \ \theta(z)
\end{equation}
where $C_n$ normalizes the probability
to find a neutron anywhere to $1$. The sequence of the Airy
function zeros has no simple analytic expression, but using the
Bohr-Sommerfield rules, we find a fairly good approximation of
this sequence:
\begin{equation}
\lambda_n \approx \left( \frac{3 \pi}{8}(4n-1) \right)^{2/3},
\end{equation}
this approximation is known to be very good even for the lowest
states.


In order to evaluate the rate for a bouncer to make a transition
$k \rightarrow n$, we will follow the semi-classical analysis as
in ref. \cite{Weinberg}. This procedure consists in deriving the
classical gravitational power $P$ emitted by an oscillating
quadrupole $Q \cos(\omega t)$, and by replacing the classical
quadrupole by the quantum quadrupole moment for the transition $k
\rightarrow n$ which reads
\begin{equation}
Q_{k n} = m \ \langle k | \hat{z}^2 | n \rangle,
\end{equation}
The quantum mechanical transition rate is:
\begin{equation}
\Gamma^{sp}_{k \rightarrow n} = \frac{P}{\hbar \omega_{k n}} =
\frac{4}{15} \frac{\omega_{k n}^5}{M_{Pl}^2 c^4} \ Q_{k n}^2,
\label{gamma}
\end{equation}
$M_{Pl}$ is the Planck mass and $\omega_{k n} = (E_k - E_n)/\hbar$ is the angular frequency of the transition. This formula is valid if the
quadrupole approximation $\omega_{k n} \ z_k \ll c$ holds, i.e. $k
\ll 10^{8}$.

This semi-classical derivation for spontaneous emission of
gravitons was recently considered in ref.~\cite{Boughn:2006st} for
atomic hydrogen. The authors provide a field theory derivation of
Eq. (\ref{gamma}) and show that it satisfies the detailed balance.

Since the quadrupole matrix elements for the quantum bouncer are
known explicitly in terms of the Airy function zeros
\cite{Goodmanson:2000}, we get:
\begin{equation}
\langle k | \hat{z}^2 | n \rangle = \frac{24
(-1)^{k-n+1}}{(\lambda_k - \lambda_n)^4} z_0^2.
\end{equation}
Or, the transition probability is equal to
\begin{equation}
\Gamma^{sp}_{k \rightarrow n} = \frac{512}{5} \frac{1}{(\lambda_k
- \lambda_n)^3} \left( \frac{m}{M_{Pl}} \right)^2 \frac{E_0^5 \
z_0^4 \ c}{(\hbar c)^5} = \frac{5 \times 10^{-77} \rm{
s}^{-1}}{(\lambda_k - \lambda_{n})^3}.
\end{equation}
For the two lowest quantum states $\lambda_2 - \lambda_1 = 1,75$
and the probability of the spontaneous graviton emission is as low
as:
\begin{equation}
\Gamma^{sp}_{2 \rightarrow 1} \sim 10^{-77} \rm{ s}^{-1}.
\end{equation}


\section*{References}
\begin{thebibliography}{10}

\bibitem{Nesvizhevsky:2002ef}
  V.~V.~Nesvizhevsky {\it et al.},
  Nature {\bf 415}, 297 (2002).

\bibitem{Nesvizhevsky:2003ww}
  V.~V.~Nesvizhevsky {\it et al.},
  Phys.\ Rev.\ D {\bf 67}, 102002 (2003)

\bibitem{Nesvizhevsky:2005}
  V.~V.~Nesvizhevsky {\it et al.},
  J. Phys.\ C {\bf 67},  (2005)

\bibitem{Rothman:2006fp}
  T.~Rothman and S.~Boughn,
  Foundations of Physics {\bf 36},  (2006) 1801.

\bibitem{Weinberg}
  S.~Weinberg,
  ``Gravitation and cosmology'' (John Wiley, New York, 1972), chap. 10.

\bibitem{Boughn:2006st}
  S.~Boughn and T.~Rothman,
  Class.\ Quant.\ Grav.\  {\bf 23}, 5839 (2006)

\bibitem{Goodmanson:2000}
  D.~M.~Goodmanson,
  Am.\ J.\ Phys.\ {\bf 68}, 866–868 (2000).



\end{thebibliography}
\end{document}